\documentclass[letterpaper,twocolumn,prx,aps,showpacs,superscriptaddress,floatfix]{revtex4-1}
\usepackage[latin1]{inputenc}
\usepackage{bm}
\usepackage[usenames]{color}
\usepackage{multirow}
\usepackage{amssymb}
\usepackage{amsbsy}
\usepackage{amsmath}
\usepackage{stmaryrd}
\usepackage{graphicx}
\usepackage{epsfig}
\usepackage{placeins}
\usepackage{ulem}
\usepackage{bbold}
\usepackage{braket}
\usepackage{blindtext}
\usepackage[colorlinks,linkcolor=blue,citecolor=blue,urlcolor=blue]{hyperref}

\makeatletter

\begin{document}

\title{Magnetization and energy dynamics in spin ladders: Evidence of diffusion in time, 
frequency, position, and momentum}

\author{Jonas Richter}
\email{jonasrichter@uos.de}
\affiliation{Department of Physics, University of Osnabr\"uck, D-49069 Osnabr\"uck, Germany}

\author{Fengping Jin}
\affiliation{Institute for Advanced Simulation, J\"ulich Supercomputing Centre,  
Forschungszentrum J\"ulich, D-52425 J\"ulich, Germany}

\author{Lars Knipschild}
\affiliation{Department of Physics, University of Osnabr\"uck, D-49069 Osnabr\"uck, Germany}

\author{Jacek Herbrych}
\affiliation{Department of Physics and Astronomy, The University of Tennessee, Knoxville, Tennessee 37996, USA}
\affiliation{Materials Science and Technology Division, Oak Ridge National Laboratory, Oak Ridge, Tennessee 37831, USA}

\author{Hans De Raedt}
\affiliation{Zernike Institute for Advanced Materials, University of Groningen,  
NL-9747AG Groningen, The Netherlands}

\author{Kristel Michielsen}
\affiliation{Institute for Advanced Simulation, J\"ulich Supercomputing Centre,  
Forschungszentrum J\"ulich, D-52425 J\"ulich, Germany}
\affiliation{RWTH Aachen University, D-52056 Aachen, Germany}

\author{Jochen Gemmer}
\affiliation{Department of Physics, University of Osnabr\"uck, D-49069 
Osnabr\"uck,
Germany}

\author{Robin Steinigeweg}
\email{rsteinig@uos.de}
\affiliation{Department of Physics, University of Osnabr\"uck, D-49069 Osnabr\"uck, Germany}

\date{\today}


\begin{abstract}
The dynamics of magnetization and energy densities are studied in the two-leg
spin-$1/2$ ladder. Using an efficient pure-state approach based on the concept  
of typicality, we calculate spatio-temporal correlation functions for large 
systems with up to $40$ lattice sites. In addition, two subsequent Fourier 
transforms from real to momentum space as well as from time to frequency domain 
yield the respective dynamical structure factors. Summarizing our main results, 
we unveil the existence of genuine diffusion both for spin and energy. In 
particular, this finding is based on four distinct signatures which can all be 
equally well detected: (i) Gaussian density profiles, (ii) time-independent 
diffusion coefficients, (iii) exponentially decaying density modes, and (iv) 
Lorentzian line shapes of the dynamical structure factor. The combination of (i) 
- (iv) provides a comprehensive picture of high-temperature dynamics in this
archetypal nonintegrable quantum model.
\end{abstract}

\maketitle


\section{Introduction}

The study of low-dimensional spin systems is one of the most active fields in 
condensed matter physics. On the one hand, quantum spin models are of immediate 
relevance to describe the properties of various Mott insulators, where (quasi) 
one-dimensional structures like chains or ladders are realized within the bulk 
materials. The notion of property can be manifold in this context, including 
thermodynamic quantities \cite{johnston2000}, transport characteristics such as 
the heat conductivity \cite{Sologubenko2000, Hess2001}, as well as other 
dynamic features probed by, e.g., neutron scattering \cite{Lake2013, 
Mourigal2013}, NMR \cite{Thurber2001}, and $\mu$SR \cite{Maeter2013}, to name 
just a few. Particularly, developing a thorough understanding of quantum 
magnets, both experimentally and theoretically, is also of essential importance 
in order to pave the way for potential spin-based technologies in the future 
\cite{Wolf2001}.  

On the other hand, from a more fundamental point of view, low-dimensional spin 
models represent prototypical examples of interacting quantum many-body 
systems, allowing to study questions ranging from the foundations of 
statistical mechanics \cite{Cazalilla2010} to the physics of black holes 
\cite{Ikhlef2012}. In particular, long-standing questions concerning the 
emergence of thermodynamic behavior in isolated quantum systems 
have recently experienced rejuvenated attention \cite{Polkovnikov2011, 
Eisert2015, Gogolin2016, Dallesio2016}. This upsurge of interest is
not least due to the advance of controlled experiments with cold atoms 
and trapped ions \cite{Langen2015, Trotzky2012, Blatt2012}, theoretical key  
concepts such as the eigenstate thermalization hypothesis \cite{deutsch1991, 
srednicki1994, rigol2005, Richter2018} and the typicality of pure quantum 
states \cite{Gemmer2004, Popescu2006, Goldstein2006, Reimann2007}, as well as 
the development of powerful numerical techniques \cite{schollwoeck20052011}. 

Concerning the relaxation in isolated quantum systems, an intriguing question  
is the difference in the properties between integrable and nonintegrable systems. On the one hand, 
integrable systems exhibit a macroscopic set of (quasi-)local conservation laws 
\cite{zotos1997, prosen2013} which might cause anomalous thermalization 
\cite{vidmar2016} as well as nondecaying currents, i.e., ballistic transport 
\cite{heidrichmeisner20032007}. Nonetheless, even for paradigmatic integrable 
models, signatures of diffusion have been reported, e.g., for the spin-$1/2$ XXZ 
chain above the isotropic point \cite{Sirker2011, znidaric2011, 
steinigeweg2011_1, karrasch2014_2, Steinigeweg2017, Ljubotina2017} or the 
Fermi-Hubbard model for strong particle-particle interactions \cite{prosen2012,  
karrasch2017, steinigeweg2017_2}. Moreover, significant progress in 
understanding transport in integrable models has been recently achieved within 
the framework of generalized hydrodynamics \cite{Alvaredo2016, Bertini2016, 
DeNardis2018}. On the other hand, in more realistic situations, integrability 
is often lifted due to various perturbations, e.g., spin-phonon coupling 
\cite{Chernyshev2016}, long-range interactions \cite{Hazzard2014}, 
dimerization \cite{Karrasch2013}, as well as the presence of impurities 
\cite{Metavitsiadis2010} or disorder \cite{Herbrych2013}. Such nonintegrable 
systems are commonly expected to have vanishing Drude weights 
\cite{heidrichmeisner20032007} and potentially exhibit diffusive transport, 
e.g., due to quantum chaos. Notably, the onset of diffusive hydrodynamics under 
chaotic quantum dynamics has also been substantiated by exact results obtained 
in random circuit models \cite{Keyserlingk2018, Nahum2018, Khemani2018}.
However, since the dynamics of real interacting systems with many degrees 
of freedom poses a formidable challenge to theory and numerics, signatures of 
clean diffusion have been found only for selected examples \cite{Michel2005, 
Monasterio2005, karrasch2014_2, Medenjak2017, Richter2018_2}.

In this context, we study the dynamics of magnetization and energy in a
nonintegrable spin-$1/2$ system with ladder geometry. In particular, we will demonstrate an  
efficient numerical approach based on the concept of typicality 
\cite{Gemmer2004, Popescu2006, Goldstein2006, Reimann2007, Hams2000, iitaka2003, 
sugiura2013, elsayed2013, monnai2014, steinigeweg2014}, which enables us to 
study large system sizes and to detect various signatures of diffusive 
transport.

This paper is structured as follows. We introduce the model in 
Sec.~\ref{Sec::Model} and define the observables in 
Sec.~\ref{Sec::Observ}. In Sec.~\ref{Sec::DQT}, we outline our numerical 
approach and present our results in Sec.~\ref{Sec::Results}. We summarize and 
conclude in Sec.~\ref{Sec::Con}.

\section{The model}\label{Sec::Model}
 
The Hamiltonian $\mathcal{H} = \sum_{l=1}^L h_l$ of the spin-$1/2$ ladder with 
periodic boundary conditions reads  
\begin{equation}\label{Eq::Hamiltonian}
h_l = J_{||} \sum_{k = 1}^2 {\bf S}_{l,k} \cdot {\bf S}_{l+1,k} + 
\frac{J_\perp}{2}\sum_{n=l}^{l+1} {\bf S}_{n,1} \cdot {\bf S}_{n,2}\ , 
\end{equation}
where ${\bf S}_{\bf r} = (S_{\bf r}^x,S_{\bf r}^y,S_{\bf r}^z)$ are spin-$1/2$  
operators on lattice site ${\bf r} = (l,k)$, $J_{||} = 1$ denotes the coupling 
along the legs (and sets the energy scale throughout this paper), and $J_\perp 
\geq 0$ is the coupling on the $L$ rungs. While for $J_\perp = 0$, 
$\mathcal{H}$ decouples into two separate chains and is integrable in terms of 
the Bethe ansatz \cite{kluemper2002}, this integrability is broken for any  
$J_\perp \neq 0$. Numerous works \cite{Zotos2004, Jung2006, Langer2009, 
znidaric2013_1, Karrasch2015, Steinigeweg2016} have explored the dynamics of 
the spin ladder \eqref{Eq::Hamiltonian}, also including various modifications 
such as four-spin terms \cite{Nishimoto2009}, Kitaev-type couplings 
\cite{Metavitsiadis2017}, as well as XX-ladder systems \cite{znidaric2013_2, 
steinigeweg2014_2}. However, while these studies often discuss either spin {\it 
or} energy transport only, they also mostly focus exclusively on the dynamics 
of densities {\it or} currents, either in time {\it or} in frequency.

In this paper, we do not focus on a particular quantity and representation and 
provide a comprehensive picture of high-temperature dynamics in the spin-$1/2$ 
ladder. As a main result, we unveil the existence of genuine diffusion {\it 
both} for magnetization and energy. In particular, this result is based on {\it 
the combination of four distinct signatures}: (i) Gaussian density profiles, 
(ii) time-independent diffusion coefficients, (iii) exponentially decaying 
density modes, and (iv) Lorentzian line shapes of the dynamical structure factor.
We present these signatures for large systems with up to $40$ lattice sites. 
\begin{figure}[tb]
\centering
\includegraphics[width=0.9\columnwidth]{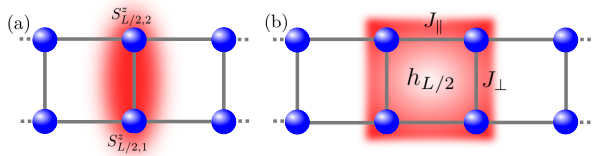}
\caption{(Color online) Sketch of the local densities $\varrho_l$. (a) local 
magnetization $S_{l,1}^z + S_{l,2}^z$. (b) local energy $h_l$. Note that $h_l$ 
is defined with $J_\perp/2$, cf.\ Eq.\ \eqref{Eq::Hamiltonian}.}
\label{Fig1}
\end{figure}
%


\section{Setup and observables}\label{Sec::Observ}

We study the dynamics of time-dependent expectation values
\begin{equation}\label{Eq::ExpV}
 p_l(t) = \bra{\psi(t)}\varrho_l \ket{\psi(t)}\ ;\quad  \varrho_l = \begin{cases}
S_{l,1}^z + S_{l,2}^z \\
h_l 
\end{cases}\ ,
\end{equation} 
where the time argument has to be understood as $\ket{\psi(t)} =  
e^{-i\mathcal{H}t} \ket{\psi(0)}$, and the operator $\varrho_l$ denotes the 
local densities of magnetization or energy (cf.\ Fig.\ \ref{Fig1}).
Furthermore, the (unnormalized) pure initial state $\ket{\psi(0)}$ is prepared as
\begin{equation}\label{Eq::State1}
\ket{\psi(0)} = 
\frac{\sqrt{\tilde{\varrho}_{L/2}}\ket{\varphi}}{\sqrt{\braket{\varphi|\varphi} 
} }\ ;\quad \ket{\varphi} = \sum_{k=1}^{d} c_k \ket{\varphi_k}\ , 
\end{equation}
where the complex coefficients $c_k$ are randomly drawn from a Gaussian  
distribution with zero mean (Haar measure \cite{bartsch2009}) and the 
$\ket{\varphi_k}$ denote a set of orthonormal basis states (e.g.\ the Ising 
basis) of the full Hilbert space with dimension $d=4^L$. Moreover, the operator 
$\tilde{\varrho}_{L/2} = \varrho_{L/2} + c$ in Eq.\ \eqref{Eq::State1} is 
essentially equivalent to $\varrho_{L/2}$ except for a constant offset
which renders the eigenvalues of $\tilde{\varrho}_{L/2}$ nonnegative. 
Exploiting the concept of dynamical typicality \cite{bartsch2009, Reimann2018}, 
as well as $\text{Tr}[\varrho_l] = 0$, the expectation value $p_l(t)$ can be 
connected to an equilibrium correlation function at formally infinite 
temperature (see Sec.~\ref{Sec::DQT}),
\begin{equation}\label{Eq::Typ}
p_l(t) = \langle \varrho_l(t) \varrho_{L/2} \rangle + \epsilon\ , 
\end{equation}
with $\langle \cdot \rangle = \text{Tr}[\cdot]/d$. Note that the statistical 
error $\epsilon=\epsilon(\ket{\varphi})$ scales as $\epsilon \propto 
1/\sqrt{d}$ and is negligibly small for all system sizes studied here 
\cite{Hams2000, sugiura2013, bartsch2009, Reimann2018}. 

In addition to the spatio-temporal correlation functions \eqref{Eq::Typ}, 
the respective correlations in momentum space can be obtained by a lattice 
Fourier transform \cite{Fabricius1997}
\begin{equation}\label{Eq::Fourier1}
 p_q(t) = \sum_l e^{iq(l-L/2)} p_l(t) = \langle \varrho_q(t) \varrho_{-q}\rangle\ , 
\end{equation}
where translational invariance has been exploited, and $\varrho_q = 
\sqrt{1/L}\sum_l  e^{iql} \varrho_l$ with discrete momenta $q = 2\pi k/L$ and $k 
= 0,1,\dots,L-1$. Furthermore, a subsequent Fourier transform from time to 
frequency domain eventually yields the dynamical structure factor $S(q,\omega)$,
\begin{equation}\label{Eq::FourierOm}
 p_q(\omega) = \int_{-t_\text{max}}^{t_\text{max}} e^{i\omega t} p_q(t)\  
\text{d}t = S(q,\omega)\ , 
\end{equation}
with the finite frequency resolution $\delta \omega = \pi/t_\text{max}$.

It is further instructive to establish a relation between density dynamics and 
current-current correlation functions. To this end, let us introduce the 
time-dependent diffusion coefficient \cite{Steinigeweg2009}
\begin{equation}\label{Eq::Dt}
D(t) = \frac{1}{\chi} \int_0^t \frac{\langle j(t') j \rangle}{L}\  \text{d}t'\ ,
\end{equation}
where $\chi = \lim_{q\to 0} \langle \varrho_{q} \varrho_{-q} \rangle$ denotes 
the isothermal susceptibility \cite{Suscep} and the spin- or energy-current 
operators $j=\sum_l j_l$ follow from a lattice continuity equation $\partial_t 
\varrho_l = i[\mathcal{H},\varrho_l] = j_{l-1} - j_{l}$. More details on current 
operators and autocorrelation functions are provided in Appendix~\ref{Sec::Cur}.

To proceed, we note that the states $\ket{\psi(0)}$ realize an initial density  
profile $p_l(0)$ which exhibits a $\delta$ peak at $l=L/2$ \cite{InitDelta}.
This initial $\delta$ peak will gradually broaden with time and its spatial  
variance for $t \geq 0$ is given by
\begin{equation}\label{Eq::Sigma}
\sigma(t)^2 = \sum_{l=1}^L l^2 \delta p_l(t) - \left(\sum_{l=1}^L l \delta 
p_l(t) \right)^2\ ,  
\end{equation}
with $\delta p_l(t) = p_l(t)/\sum_l p_l(0)$ and $\sum_l \delta p_l(t) = 1$. Due 
to the typicality relation in Eq.\ \eqref{Eq::Typ}, this variance can be 
directly connected to the aforementioned diffusion coefficient 
\cite{steinigeweg2009_2, yan2015}, 
\begin{equation}\label{Eq::Sigt}
\frac{\text{d}}{\text{d}t}\sigma(t)^2 = 2D(t)\ . 
\end{equation}
Note that a diffusive process requires $D(t) = D = \text{const}$.\ such that  
$\sigma(t) \propto \sqrt{t}$ \cite{karrasch2014_2}, above the mean free time.


\section{Numerical Approach}\label{Sec::DQT}

\subsection{Dynamical quantum typicality}

First, let us derive the typicality relation given in Eq.\ \eqref{Eq::Typ}. 
To this end, we start with an infinite-temperature correlation function, 
\begin{align}
{\cal C}_{l,l'}(t) &= \frac{\text{Tr}[\varrho_l(t) \tilde{\varrho}_{l'}]}{d}
= \frac{\text{Tr}[\varrho_l(t) \varrho_{l^\prime}]+c\text{Tr}[\varrho_l]}{d}\ 
\label{Eq::Eq2} ,
\end{align}
where $\tilde{\varrho}_{l'} = \varrho_{l'}+c$ has a nonnegative spectrum. We 
realize that Eq.\ \eqref{Eq::Eq2} can be simplified if either $c = 0$ or 
$\text{Tr}[\varrho_l] = 0$. Focusing on these cases, one finds 
\begin{equation}\label{Eq::Eq3}
{\cal C}_{l,l'}(t) = \frac{\text{Tr}[\varrho_l(t) \varrho_{l^\prime}]}{d} = 
\langle 
\varrho_l(t) \varrho_{l^\prime} \rangle\ ,  
\end{equation}
i.e., the correlation functions $\langle \varrho_l(t) \tilde{\varrho}_{l'}  
\rangle$ and $\langle \varrho_l(t) \varrho_{l^\prime} \rangle$ are 
equivalent. This is a first important observation. Furthermore, exploiting the 
cyclic invariance of the trace, Eq.~\eqref{Eq::Eq2} can be written as  
\begin{equation}\label{Eq::Eq4}
{\cal C}_{l,l'}(t) = 
\frac{\text{Tr}[\sqrt{\tilde{\varrho}_{l'}}\varrho_l(t)\sqrt{\tilde{\varrho}_{l'
}}]}{d}\ ,
\end{equation}
where the square root operation has to be understood in a representation where 
$\tilde{\varrho}_{l'}$ is diagonal (cf.~Sec.~\ref{Sec::NumAp}). Using the 
concept of quantum typicality \cite{Gemmer2004, Popescu2006, Goldstein2006, 
Reimann2007, Hams2000, iitaka2003, sugiura2013, elsayed2013, monnai2014, 
steinigeweg2014}, the trace in Eq.\ \eqref{Eq::Eq4} can be replaced by a scalar 
product with a single pure state $\ket{\varphi}$ which is randomly drawn 
according to the unitary invariant Haar measure,  
\begin{align}
{\cal C}_{l,l'}(t) &\approx 
\frac{\bra{\varphi}\sqrt{\tilde{\varrho}_{l'}}\varrho_l(t)\sqrt{\tilde{\varrho}_
{l'}}\ket{\varphi}}{\braket{\varphi|\varphi}} + \epsilon(\ket{\varphi}) 
\label{Eq::Eq5}\\
&= \bra{\psi(0)} \varrho_l(t) \ket{\psi(0)} = \bra{\psi(t)} \varrho_l 
\ket{\psi(t)}\label{Eq::Eq6}\ , 
\end{align}
where we have used the definition of $\ket{\psi(0)}$ from Eq.\ 
\eqref{Eq::State1}  and interpreted the time dependence as a 
property of the states and not of the operator. The combination of 
Eqs.~\eqref{Eq::Eq2} to \eqref{Eq::Eq6} yields the typicality relation 
\eqref{Eq::Typ}, where we have chosen $l' = L/2$ without loss of generality.  

\subsection{Construction of initial states}\label{Sec::NumAp}

Concerning the construction of the pure state $\ket{\psi(0)}$ in Eq.~\eqref{Eq::State1}, i.e., the evaluation of the square root $\sqrt{\tilde{\varrho}_{L/2}}$, 
the following comments are in order. On the one hand, in the case of $\varrho_l = S_{l,1}^z + S_{l,2}^z$ this procedure is 
rather simple since $S_{l,1}^z + S_{l,2}^z$ is naturally diagonal in the Ising basis, which is routinely used as our working basis. 
On the other hand, in the case of the local energy, $\varrho_l = h_l$ is 
not diagonal immediately. While this situation usually requires diagonalization, a complete diagonalization of $h_l$ can 
still be avoided since $h_l$ is a local operator acting nontrivially only on a
small part of the product space. Thus, although the preparation of 
$\ket{\psi(0)}$ becomes more 
demanding in the case of $\varrho_l = h_l$, it certainly remains feasible and yields a powerful numerical approach as well.  
If one still wants to refrain from such square-root constructions, it is also possible to use \textit{two} auxiliary pure states instead of just \textit{one} 
(cf.\ Appendix~\ref{Sec::Cur}). It should be noted however, that the approach presented in this paper, using just a single pure state, 
will generally be more favorable concerning memory requirements and run time 
(even if the initial preparation of $\ket{\psi(0)}$ is more costly).

\subsection{Pure-state propagation}

Relying on the typicality relation \eqref{Eq::Typ}, we calculate spatio-temporal 
correlation functions for spin and energy densities. The main advantage of this approach comes from the fact that 
the action of $e^{-i\mathcal{H}t}$ on the pure state $\ket{\psi(0)}$ can be 
efficiently evaluated {\it without} the diagonalization of $\mathcal{H}$, e.g., 
by means of a Trotter decomposition \cite{deReadt2006} (see also Appendix \ref{Sec::Trotter}), or also by other approaches \cite{Dobrovitski2003, Weisse2006, Varma2017}.
Moreover, let us stress that the numerical costs of the Fourier transforms \eqref{Eq::Fourier1} 
and \eqref{Eq::FourierOm} are practically negligible. Therefore, we essentially obtain 
all information on the dynamics of either magnetization or energy 
from the time evolution of the {\it single} pure state $\ket{\psi(t)}$ and the measurement of $L$ local 
operators $\varrho_l$, cf.\ Appendix~\ref{Sec::FT}. 
In practice, this pure-state approach enables us to treat ladders with up to $L = 20$ rungs, i.e., $40$ lattice sites in total. 
If not stated otherwise, we always take into account the full Hilbert space (e.g.\ $d\approx 10^{12}$ for $L = 20$).

\subsection{Finite temperatures}

Although the focus of this paper is on quantum dynamics at formally infinite 
temperature, let us briefly outline how finite-temperature correlations can be 
obtained based on pure-state calculations as well. On the one hand, a 
straightforward approach is the construction of the typical pure state 
$\ket{\varphi_\beta}$ according to \cite{sugiura2013, steinigeweg2014}
\begin{equation}
\ket{\varphi_\beta} = e^{-\beta {\cal H}/2} \ket{\varphi}\ ,
\end{equation}
where the reference pure state $\ket{\varphi}$ corresponds to infinite 
temperature and has been introduced in Eq.\ \eqref{Eq::State1}. Analogous to 
the real-time evolution, the action of $e^{-\beta {\cal H}/2}$ can be 
efficiently evaluated by an iterative forward propagation, but now in imaginary 
time. More details on finite-temperature calculations using 
$\ket{\varphi_\beta}$ can be found in Appendix~\ref{Sec::Cur}.

On the other hand, another useful class of pure states 
$\ket{\tilde{\varphi}_{\alpha,\beta}}$, which has been put forward in 
Ref.~\cite{Richter2019}, is constructed as
\begin{equation}
 \ket{\tilde{\varphi}_{\alpha,\beta}} \propto e^{-\beta({\cal H} - \alpha \varrho_{l'})/2} \ket{\varphi}\ . 
\end{equation}
In particular, depending on the size of the control parameter $\alpha$, these 
states can not only be used to obtain linear-response correlation functions, 
but also to calculate far-from-equilibrium quantum dynamics \cite{Richter2018, 
Richter2019}.

Finally, let us note that, since the effective Hilbert-space dimension shrinks 
for $\beta > 0$, the statistical error $\epsilon$ of the typicality 
approximation, cf.\ Eq.\ \eqref{Eq::Typ}, is generally larger compared to the 
infinite-temperature limit. Nevertheless, even for $\beta > 0$ this error still 
decreases exponentially with system size and accurate calculations remain 
possible for moderate temperatures.


%
\begin{figure}[tb]
 \centering
 \includegraphics[width=0.9\columnwidth]{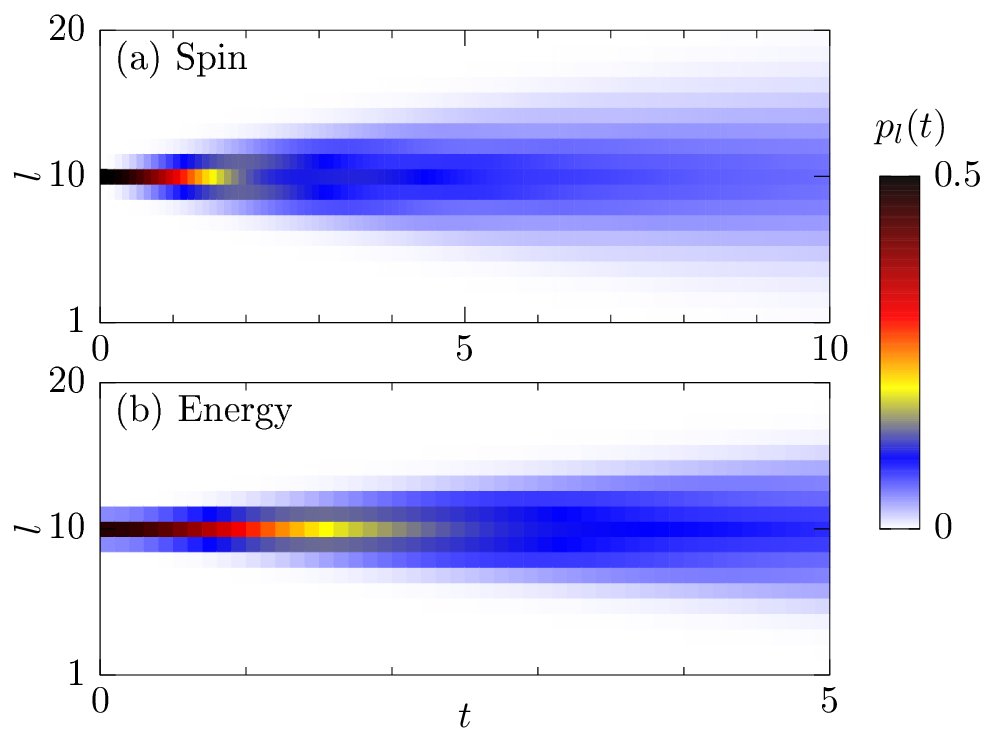}
 \caption{(Color online) Real-time broadening of density profiles 
 $p_l(t)$ for spin $\varrho_l = S_{l,1}^z + S_{l,2}^z$ and energy $\varrho_l = h_l$. 
 The other parameters are $J_\parallel = J_\perp = 1$ and $L = 20$.}
 \label{Fig2}
\end{figure}

\section{Results}\label{Sec::Results}

We now present our numerical results. In Sec.~\ref{Sec::RSD}, we start with the analysis of density dynamics in real space. 
The corresponding structure factors are then discussed for magnetization in Sec.~\ref{Sec::SFS} and for energy in Sec.~\ref{Sec::SFE}.

\subsection{Real-space dynamics}\label{Sec::RSD}

To begin with, let us focus on the isotropic case $J_\parallel = J_\perp = 1$. 
Starting with dynamics in time and real space, Figs.\ \ref{Fig2} (a) and (b) show the density profiles $p_l(t)$ 
of magnetization and energy for large systems with $L = 20$. 
One can clearly observe the initial $\delta$ peak at $t = 0$ (or almost $\delta$ peak \cite{InitDelta}) which 
broadens for times $t > 0$. Moreover, on the (short) time scales depicted, $p_l(t)$ does not 
reach the boundaries of the system, i.e., trivial finite-size effects do not occur. 
\begin{figure}[tb]
 \centering
 \includegraphics[width=\columnwidth]{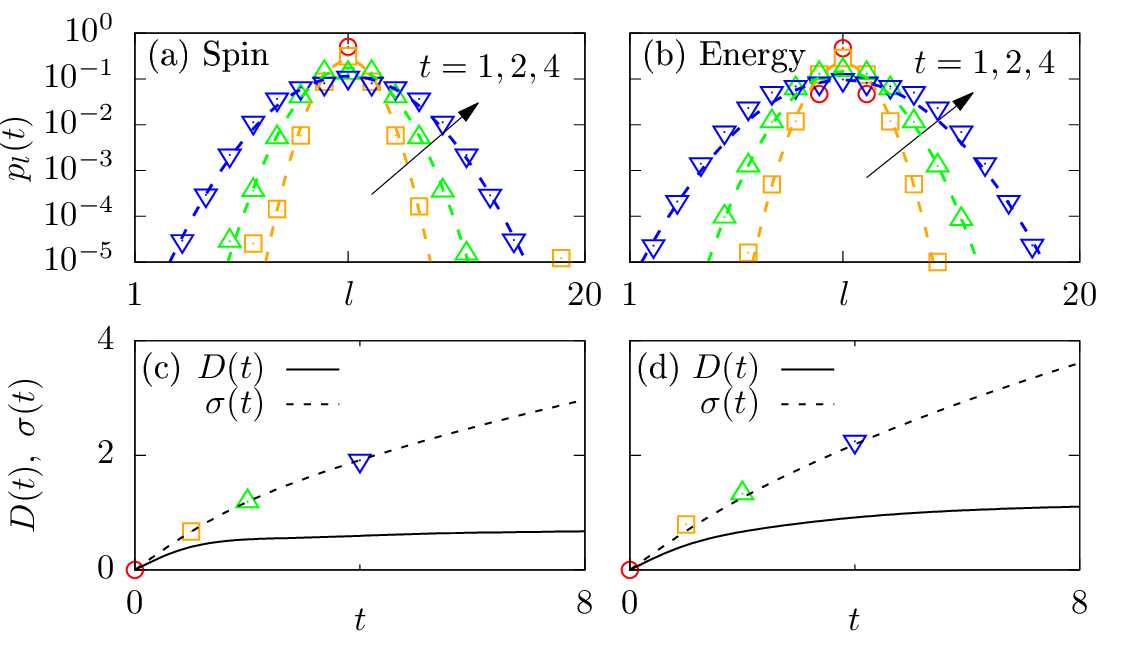}
 \caption{(Color online) (a) and (b): Density profiles $p_l(t)$ for spin and energy, 
 at fixed times $t= 0$ ($\delta$ peak) and $t = 1,2,4$ (arrow). The dashed lines 
 are Gaussian fits to the data. (c) and (d): Widths of the density profiles (symbols) obtained from Eq.\ \eqref{Eq::Sigma},
 as well as $D(t)$ and $\sigma(t)$ (lines) calculated from current autocorrelations, cf.\ Eqs.\ \eqref{Eq::Dt} and \eqref{Eq::Sigt}. For sufficiently 
 long times, one finds $D(t) \approx \text{const}$. 
 The other parameters are $J_\parallel = J_\perp = 1$, $L = 20$ (densities), $L = 18$ (spin current), and $L = 15$ (energy current \cite{Steinigeweg2016}).}
 \label{Fig3}
\end{figure}

For a more detailed discussion, Figs.\ \ref{Fig3} (a) and (b) show 
cuts of $p_l(t)$ at fixed times $t = 0, 1, 2, 4$. For these times, 
one finds that the data are well described by 
Gaussians over several orders of magnitude,
\begin{equation}\label{Eq::Gauss}
 p_l(t) \propto \exp\left[-\frac{(l-L/2)^2}{2\sigma(t)^2}\right]\ . 
\end{equation}
While these Gaussians already suggest diffusion both for magnetization 
and energy (cf.~Appendix~\ref{Sec::Diff}), it is only an sufficient criterion if $\sigma(t)$ scales as $\sigma(t) \propto \sqrt{t}$ as well. 
Consequently, Figs.\ \ref{Fig3} (c) and (d) show the widths of the density profiles obtained from Eq.\ \eqref{Eq::Sigma} 
(symbols) in comparison with the respective quantities $D(t)$ and $\sigma(t)$ (lines), 
calculated from the current autocorrelations, cf.\ Eqs.\ \eqref{Eq::Dt} and \eqref{Eq::Sigt}. Generally, one observes an excellent agreement between density 
and current dynamics. Moreover, after a linear increase at short times, 
$D(t)$ eventually saturates at a constant plateau $D(t) \approx \text{const.}$, and correspondingly $\sigma(t) \propto \sqrt{t}$. 
Thus, based on our numerical analysis in time and real space, we unveil the existence of diffusive 
transport in the spin-$1/2$ ladder both for magnetization as well as energy. 
This is a first main result of this paper. 

\begin{figure}[tb]
 \centering
 \includegraphics[width=0.8\columnwidth]{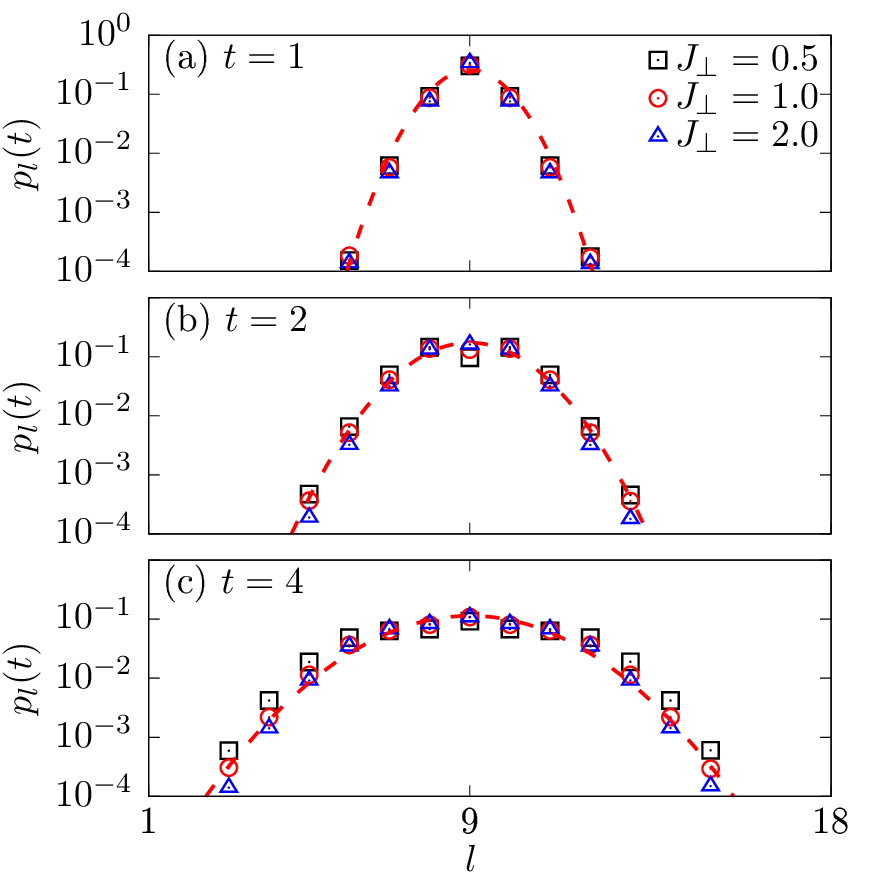}
 \caption{(Color online) Magnetization profiles $p_l(t)$ at fixed times $t = 1,2,4$. 
 Comparison between different interchain couplings $J_\perp = 0.5,1,2$. The dashed lines are Gaussian fits to the data. The 
 other parameters are $J_\parallel = 1$ and $L = 18$.}
 \label{Fig4}
\end{figure}

Let us now briefly present magnetization profiles also for other ratios $J_\perp/J_\parallel \neq 1$. 
To this end, Fig.\ \ref{Fig4} shows the real-space density profiles $p_l(t)$ at fixed times $t = 1,2,4$ for interchain couplings 
$J_\perp = 0.5,1,2$ ($J_\parallel = 1$). Generally, we find that the profiles are very similar to each other for all $t$ and
$J_\perp$ shown here. In particular, all profiles are convincingly described by Gaussian fits over several orders of magnitude. 

\subsection{Spin structure factor}\label{Sec::SFS}

Next, let us also study magnetization dynamics in momentum space, 
where the lattice diffusion equation decouples into separate Fourier modes (cf.~Appendix~\ref{Sec::Diff}). 
 
\subsubsection{Long-wavelength limit}

\begin{figure}[tb]
 \centering
 \includegraphics[width=0.85\columnwidth]{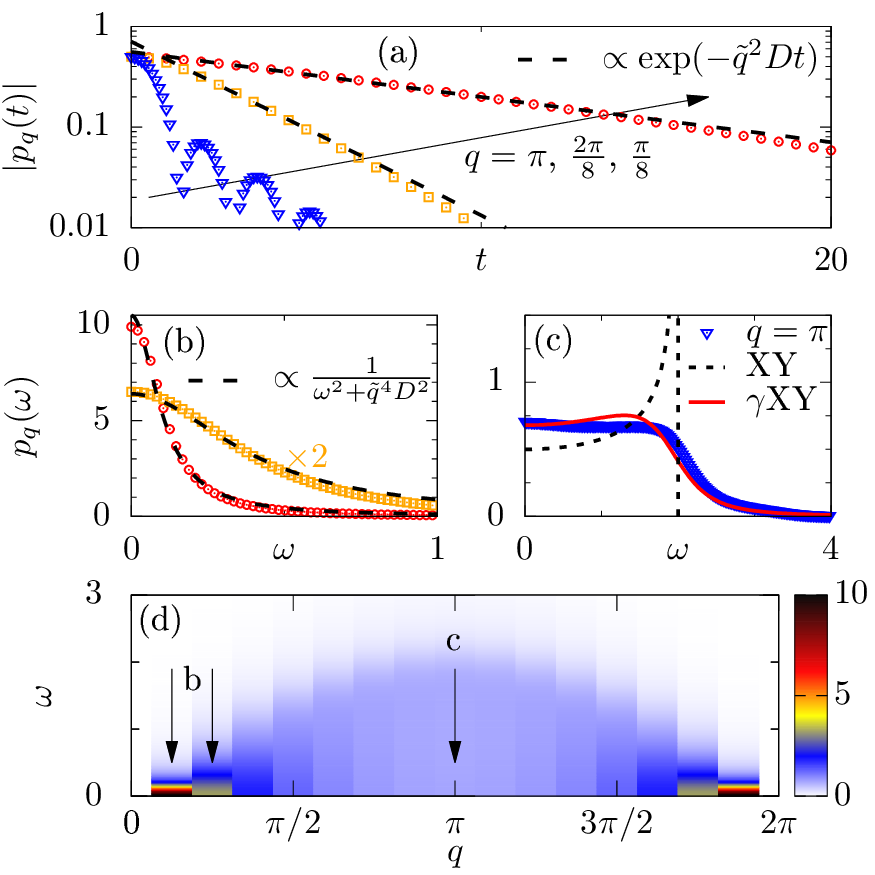}
 \caption{(Color online) Spin dynamics. (a): $p_q(t)$ for wave numbers $k = 1,2,8$ in a semilogarithmic plot. 
 Lines are exponential functions according to Eq.\ \eqref{Eq::ExpLor}. (b): $p_q(\omega)$ for $k = 1, 2$. Lines are Lorentzians according to Eq.\ \eqref{Eq::ExpLor2}. 
 (c): $p_q(\omega)$ for $k = 8$. 
 Comparison with XY model [Eq.\ \eqref{Eq::XY}] and an effective model ($\gamma$XY) where the original 
 memory kernel of the Bessel function is exponentially damped (here $\gamma = 0.6$) \cite{Knippschild2018}. 
 The XY data is multiplied by a factor of 2 in order to account for the two legs of the ladder.
(d): $p_q(\omega)$ in the full Brillouin zone. Arrows show the data of panels 
 (b) and (c). Other parameters: $J_\parallel = J_\perp = 1$, $L = 16$, and $\delta \omega = \pi/150$.} 
 \label{Fig5}
\end{figure}

In Fig.\ \ref{Fig5} (a), the density modes $p_q(t)$ for $J_\parallel = J_\perp = 1$ are shown for various momenta $q$ in a semilogarithmic plot. 
On the one hand, for large $q = \pi$, $p_q(t)$ exhibits pronounced oscillations and 
essentially decays on a time scale $t \sim 5$. On the other hand, for the two smallest wave numbers
$k = 1$ and $k = 2$, we find clean exponential relaxation 
\begin{equation}\label{Eq::ExpLor}
 p_q(t) \propto e^{-\tilde{q}^2Dt}\ ,
\end{equation}
where we have introduced the abbreviation $\tilde{q}^2 = 2[1 - \cos(q)] \approx q^2$ for sufficiently small $q$, 
and $D = \text{const.}$ can be extracted from the constant plateau in Fig.\ \ref{Fig3} (c). 
Going from time to frequency domain, Fig.\ \ref{Fig5} (b) shows the corresponding dynamical structure factors 
$p_q(\omega)$ for $k = 1, 2$. One observes that the data can be accurately 
described by Lorentzians of the form, 
\begin{equation}\label{Eq::ExpLor2}
p_q(\omega) \propto \frac{1}{\omega^2 + \tilde{q}^4 D^2}\ .  
\end{equation}
Note that, while we display the time data in Fig.\ \ref{Fig5} (a)
only up to intermediate time scales, the Fourier transform \eqref{Eq::FourierOm} is routinely 
performed for a much longer cut-off time (here $t_\text{max} = 150$) in order to achieve a high frequency resolution.
The exponential relaxation [Fig.\ \ref{Fig5} (a)] and the Lorentzian line shapes [Fig.\ \ref{Fig5} (b)] 
clearly confirm our earlier observations in the context of 
Fig.\ \ref{Fig3}, i.e., the occurrence of genuine spin diffusion in the spin-$1/2$ ladder. This is another main result of the present work.

Eventually, Fig.\ \ref{Fig5} (d) shows the dynamical structure factor $p_q(\omega)$ for all momenta $q$. 
Let us stress again that within our numerical approach the calculation of these density modes essentially does not 
require any additional resources.  
On the one hand, for small momenta $q \to 0$, one can clearly identify the diffusion peaks discussed above.  
On the other hand, in the center of the Brillouin zone, we find that $p_q(\omega)$ exhibits a broad continuum. 
This short-wavelength limit will be discussed below in more detail. 
 
\begin{figure}[tb]
 \centering
 \includegraphics[width=0.9\columnwidth]{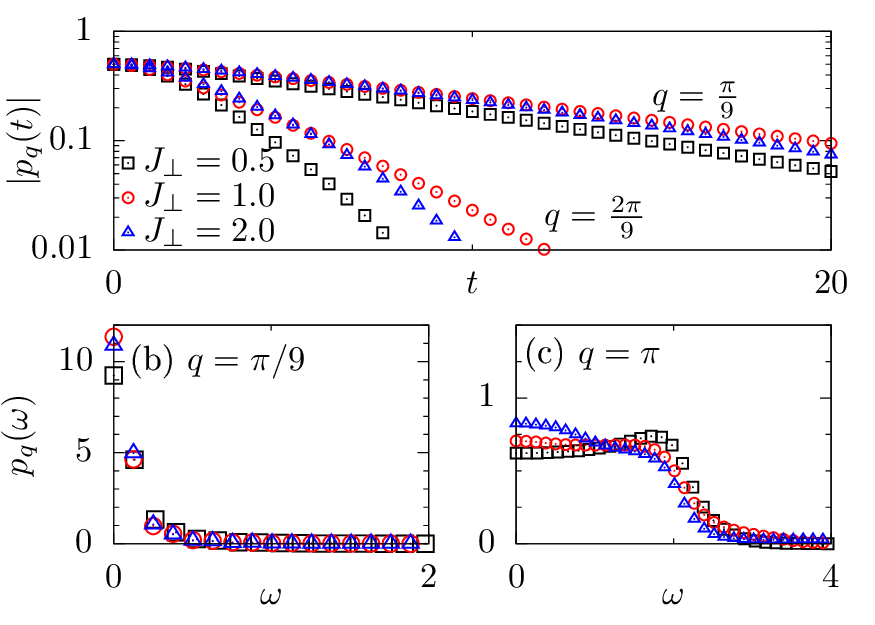}
 \caption{(Color online) Spin dynamics for different interchain couplings $J_\perp = 0.5,1,2$. (a): $p_q(t)$ for the two smallest wave numbers $k = 1,2$. 
 (b) and (c): $p_q(\omega)$ for $k = 1$ and $k = 9$, respectively. Other parameters: $J_\parallel = 1$, $L = 18$, and $\delta \omega = \pi/25$.}
 \label{Fig6}
\end{figure}

In order to corroborate once more that the emergence of diffusive transport is not restricted to the isotropic point, 
Fig.\ \ref{Fig6} shows the structure factors $p_q(t)$ and $p_q(\omega)$ for $J_\perp = 0.5, 1, 2$. 
For the smallest wave number $k = 1$, we find a clean exponential decay of $p_q(t)$ with a decay rate $-\tilde{q}^2D$, which is 
almost identical for all $J_\perp$ shown here. This fact is also reflected in the Lorentzian shape of $p_q(t)$ for 
this momentum [Fig.\ \ref{Fig6} (b)], which essentially coincides for all strengths of interchain couplings. 
On the other hand, for wave number $k = 2$, we find that $p_q(t)$ for $J_\perp = 0.5$ shows 
some deviations from an exponential, i.e., the hydrodynamic regime becomes smaller for smaller $J_\perp$, which can be 
explained by the increased mean free path of spin excitations. 

\subsubsection{Short-wavelength limit}

\begin{figure}[tb]
 \centering
 \includegraphics[width=0.8\columnwidth]{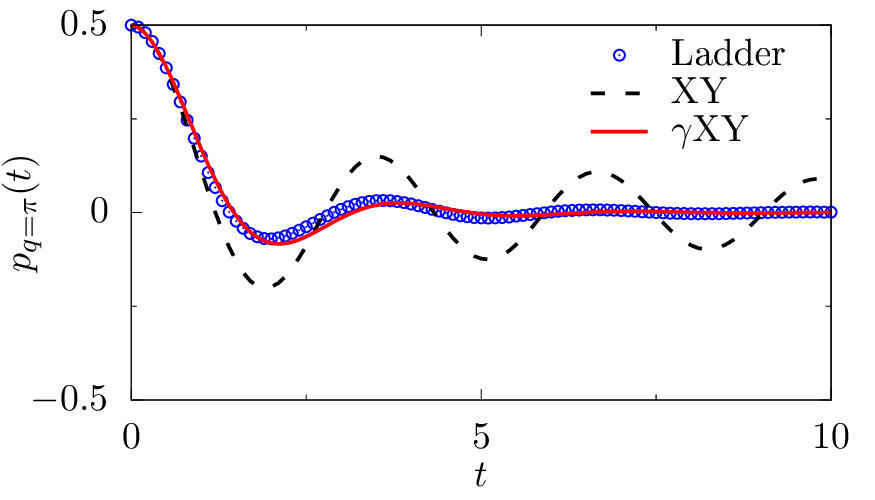}
 \caption{(Color online) Comparison of $p_{q=\pi}(t)$ between the spin ladder with $J_\parallel = J_\perp = 1$ [Fig.\ \ref{Fig5} (a) from main text], 
 the bare XY model (Bessel function), 
 and the effective dynamics $\tilde{p}_q(t)$ ($\gamma$XY) [see Eqs.\ \eqref{Eq::MemK} and \eqref{Eq::Damp}] with damping $\gamma = 0.6$. The 
 XY data is multiplied by a factor of 2 in order to account for the two legs of the ladder.}
 \label{Fig7}
\end{figure}

In addition to the long-wavelength limit, Fig.\ \ref{Fig5} (c) shows $p_{q}(\omega)$ at momentum $q = \pi$. 
For this momentum, one finds that $p_q(\omega)$ is practically $\omega$-independent up to $\omega \lesssim 2$ and exhibits a constant plateau. 
It is instructive to compare this result to the dynamics of the one-dimensional XY model.
Since the XY chain is equivalent to a model of free fermions, $p_{q}(\omega)$ is known exactly \cite{Fabricius1997}
and reads (for $\beta \to 0$, $J_\parallel = 1$, and $q = \pi$)
\begin{equation}\label{Eq::XY}
 \hspace{-0.1cm} p_{q=\pi}(\omega) = \int\limits_{-\infty}^\infty e^{i\omega t} 
 \frac{\mathcal{J}_0(2t)}{4}\ \text{d}t = 
 \frac{\Theta(2 - |\omega|)}{2\sqrt{4 - \omega^2}}\ .
\end{equation}
Here, $\mathcal{J}_0(t)$ is the Bessel function of first kind (and order zero),
and $p_{q=\pi}(\omega)$ exhibits a square-root divergence at $\omega = 2$, cf.\ Fig.\ \ref{Fig5} (c).  
Following an approach introduced in Ref.\ \cite{Knippschild2018}, the dynamics $p_q(t) \propto {\cal J}_0(2t)$ is interpreted as being generated by an 
integro-differential equation comprising a {\it memory kernel} $K(t)$, 
\begin{equation}\label{Eq::MemK}
 \frac{\text{d}}{\text{d}t}\ p_q(t) = -\int_0^t K(t-t')p_q(t') \text{d}t'\ .
 \end{equation}
Equation \eqref{Eq::MemK} establishes a direct correspondence 
between $p_q(t)$ and $K(\tau)$ and can be evaluated in both directions. Thus, given the original dynamics 
$p_q(t) \propto {\cal J}_0(2t)$, the respective memory kernel $K(\tau)$ can be calculated, e.g., numerically.
In fact, given the expression in Eq.\ \eqref{Eq::XY}, $K(\tau)$ can be even obtained analytically and reads, 
\begin{equation}
 K(\tau) = \frac{2{\cal J}_1(2\tau)}{\tau}\ . 
\end{equation}
Comparing the bare XY model with the full spin ladder \eqref{Eq::Hamiltonian}, 
the additional rung couplings as well as the $S^z_{\bf r}S^z_{\bf r'}$ terms are treated as a perturbation
giving rise to an exponential damping of this memory kernel (for small perturbations) \cite{Knippschild2018, Knipschild2019},
\begin{equation}\label{Eq::Damp}
 \tilde{K}(\tau) = K(\tau)e^{-\gamma \tau}\ . 
\end{equation}
This new memory kernel $\tilde{K}(\tau)$ is then used to numerically evaluate Eq.\ \eqref{Eq::MemK} in order to 
obtain the new (damped) dynamics $\tilde{p}_q(t)$.  
As shown in Fig.\ \ref{Fig5} (c), the effective dynamics generated by this (heuristic) approach with $\gamma = 0.6$ reproduces the structure factor 
of the spin ladder remarkably well, even though the perturbation is not small.   
Moreover, as shown in Fig.~\ref{Fig7}, this convincing agreement between spin 
ladder and effective model can not only be observed in frequency space, 
but also in real time. Eventually, let us note that, while $\gamma = 0.6$ is 
found to describe the data most accurately, 
a more quantitative understanding of this specific value goes beyond the scope of the present paper, see also 
Ref.~\cite{Knipschild2019}.

Thus, while clear signatures of 
diffusion can be found for long-wavelength modes [cf.\ Figs.\ \ref{Fig5} (a) and (b)], the relaxation of 
$p_q(t)$ for $q \to \pi$ can be qualitatively understood as the (damped) dynamics of free fermions, e.g.,
since the wavelength is smaller than the mean-free path. Note that similar behavior has been found 
also for XXZ chains~\cite{Fabricius1997, Herbrych2012}. 

\subsection{Energy structure factor}\label{Sec::SFE}

\begin{figure}[t]
 \centering
 \includegraphics[width=0.85\columnwidth]{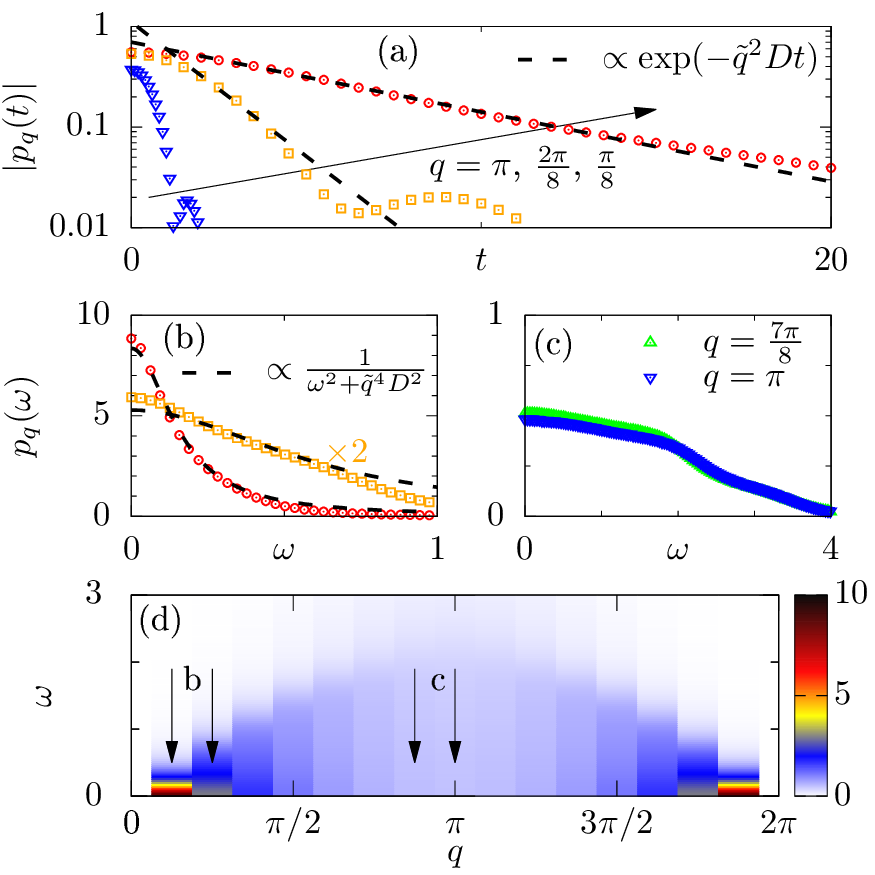}
 \caption{(Color online) Energy dynamics. (a): $p_q(t)$ for $k = 1,2,8$. (b) $p_q(\omega)$ for $k = 1,2$. 
 (c): $p_q(\omega)$ for $k = 7,8$. (d): $p_q(\omega)$ in the full Brillouin zone.  
Other parameters: $J_\parallel = J_\perp = 1$, $L = 16$, and $\delta \omega = \pi/150$}
 \label{Fig8}
\end{figure}

We now also present results for energy dynamics in momentum space. Analogously to the discussion in the context of Fig.~\ref{Fig5}, 
Fig.\ \ref{Fig8} shows the energy structure factors $p_q(t)$ and $p_q(\omega)$ of the isotropic spin ladder for short and long wavelengths. 
Generally, the results for energy dynamics are very similar to the previously discussed case of magnetization, i.e., 
$p_q(t)$ decays exponentially for small $q$ while the corresponding 
$p_q(\omega)$ exhibits a Lorentzian line shape. 
Thus, the data shown in Fig.~\ref{Fig8} confirm the existence of diffusive energy transport as well, see also Refs.~\cite{Zotos2004, Karrasch2015, Steinigeweg2016}. 

\begin{figure}[t]
 \centering
 \includegraphics[width=0.85\columnwidth]{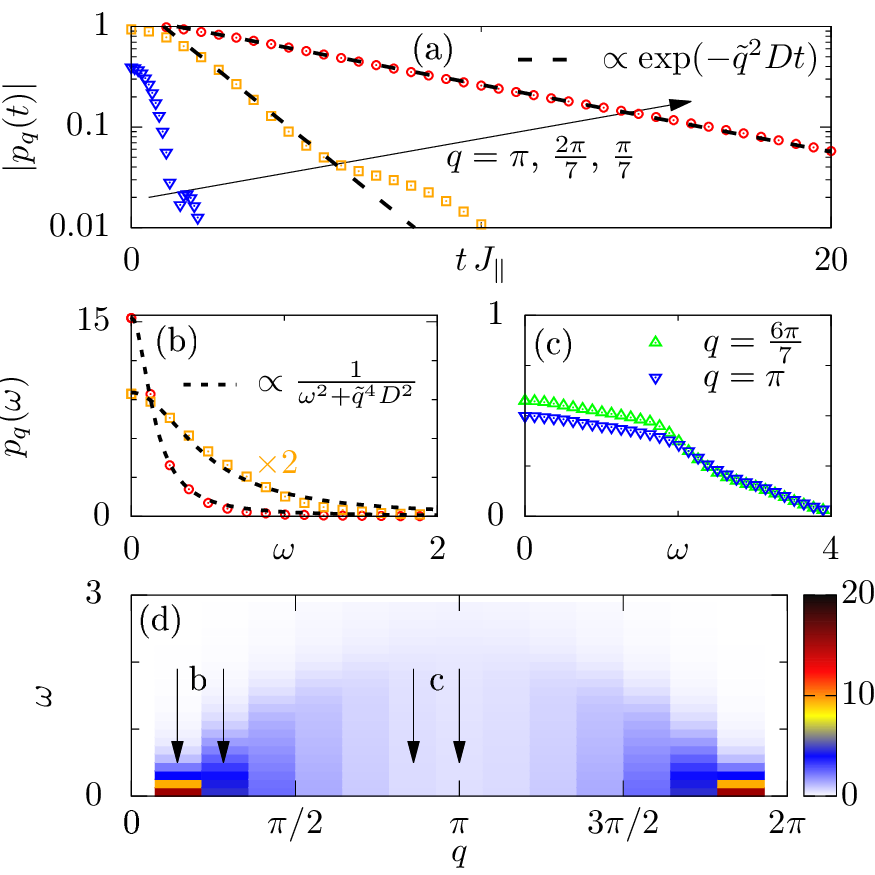}
 \caption{(Color online) Energy dynamics in the presence of a magnetic field.  
 Calculations are performed in the subsector with $S^z = 0$ only. 
 The other parameters are: $J_\parallel = J_\perp = B = 1$, $L = 14$, and $\delta \omega = \pi/25$.}
 \label{Fig9}
\end{figure}

Finally, it is instructive to discuss the effect of an additional uniform magnetic field $B > 0$ in the $z$ direction, i.e., the new local energy $\tilde{h}_l$ reads 
 \begin{equation}
  \tilde{h}_l = h_l + \frac{B}{2}\sum_{n=l}^{l+1} \sum_{k=1}^2 S_{n,k}^z\ , 
 \end{equation}
where $h_l$ is defined according to Eq.\ \eqref{Eq::Hamiltonian}. 
Such a modification results in a magnetothermal correction and the heat current takes on the form 
 \begin{equation}
  \tilde{j}_E = j_E + Bj_S\ ,
 \end{equation}
 where the spin current $j_S$ is independent of $B$.
 For specific expressions of $j_S$ and $j_E$, see Appendix~\ref{Sec::Cur}. 
In Fig.\ \ref{Fig9}, we depict the energy structure factors $p_q(t)$ and $p_q(\omega)$ in time 
as well as frequency domain for various momenta $q$. Note that we restrict ourselves to the symmetry subspace with magnetization $S^z = 0$. 
Generally, we find that the presence of a finite magnetic field 
does not qualitatively change the behavior of $p_q(t)$ and $p_q(\omega)$. Again, one can observe an exponential decay 
for the two smallest wave numbers $k = 1,2$ and correspondingly a Lorentzian line shape in frequency space at these momenta. 
Compared to the results with $B = 0$ shown in Fig.\ \ref{Fig8}, one might even argue that those signatures of 
diffusion are slightly improved due to the finite magnetic field.  


\section{Conclusion}\label{Sec::Con}

To summarize, we have studied spin and energy dynamics in the spin-$1/2$ ladder for large systems 
with up to $40$ lattice sites. 
Our state-of-the-art numerical simulations have unveiled the existence of genuine diffusion 
both for spin and energy. In particular, this finding 
is based on four distinct signatures which have all been equally well detected: 
(i) Gaussian density profiles, (ii) time-independent diffusion coefficients, 
(iii) exponentially decaying density modes, and 
(iv) Lorentzian line shapes of the dynamical structure factor. 
Combining (i) - (iv), this paper provides a comprehensive picture of high-temperature dynamics 
in the spin-$1/2$ ladder. Promising directions of research include, e.g., the application 
of the pure-state approach to a larger class of condensed matter systems and a wider range 
of temperatures \cite{Rousochatzakis2018}. In this context, it is also an 
intriguing question if the observed signatures of diffusion persist at lower 
temperatures.
%

\subsection*{Acknowledgments}

This work has been funded by the Deutsche 
Forschungsgemeinschaft (DFG) - Grants No. 397107022 (GE 1657/3-1), No. 
397067869 (STE 2243/3-1), No. 355031190 - within the DFG Research Unit FOR 2692.
Additionally, we gratefully acknowledge the computing time, granted by the 
``JARA-HPC Vergabegremium'' and provided on the ``JARA-HPC Partition'' part of 
the supercomputer ``JUWELS'' at Forschungszentrum J\"ulich. J.H. has been 
supported by the US Department of Energy (DOE), Office of Science, Basic Energy 
Sciences (BES), Materials Sciences and Engineering Division. 

\appendix

\section{Current operators and autocorrelations}\label{Sec::Cur}

The spin current $j_S$ is defined as \cite{Karrasch2015}
\begin{equation}\label{Eq::SpinCur}
 j_S = J_\parallel \sum_{l=1}^L \sum_{k=1}^2 \left (S_{l,k}^x S^y_{l+1,k} - S_{l,k}^y S^x_{l+1,k} \right)\ . 
\end{equation}
Moreover, the energy current $j_E = j_\parallel + j_\perp$ can be written as a sum of a longitudinal and a 
perpendicular part which read \cite{Zotos2004,Steinigeweg2016}
\begin{align}
 j_\parallel &= J_\parallel^2 \sum_{l=1}^L \sum_{k=1}^2 {\bf S}_{l-1,k} \cdot ({\bf S}_{l,k} \times {\bf S}_{l+1,k})\ , \label{Eq::JE1}\\
 j_\perp &= \frac{J_\parallel J_\perp}{2} \sum_{l=1}^L \sum_{k=1}^2 ({\bf S}_{l-1,k} - {\bf S}_{l+1,k}) \cdot ({\bf S}_{l,k}\times {\bf S}_{l,3-k})\ \label{Eq::JE2} .
\end{align}
In order to calculate current-current correlation functions $\langle j(t) j \rangle$ by means of a 
typicality-based approach directly (at finite or infinite temperature), we 
use the two auxiliary pure states \cite{steinigeweg2014}
\begin{align}
 \ket{\varphi(\beta,t)} &= e^{-i\mathcal{H}t} e^{-\beta \mathcal{H}/2} \ket{\varphi}\ , \label{Eq::Cur1} \\
 \ket{\phi(\beta,t)} &= e^{-i\mathcal{H}t} j e^{-\beta \mathcal{H}/2} \ket{\varphi}\ , \label{Eq::Cur2}
\end{align}
which only differ by the additional current operator in Eq.\ \eqref{Eq::Cur2}, and where 
$\ket{\varphi}$ is again a random state drawn according to the Haar measure, cf.\ Eq.\ \eqref{Eq::State1}. 
Then, we can write \cite{steinigeweg2014}
\begin{equation}\label{Eq::Cur3}
 \langle j(t) j \rangle = 
 \frac{\bra{\varphi(\beta,t)}j\ket{\phi(\beta,t)}}{\braket{\varphi(\beta,0)|\varphi(\beta,0}} + \epsilon(\ket{\varphi})\ , 
\end{equation}
where again $\epsilon(\ket{\varphi}) \propto 1/\sqrt{d}$ for $\beta \to 0$. Of course, by replacing the current 
operator $j$ in Eqs.\ \eqref{Eq::Cur2} and \eqref{Eq::Cur3}, it is straightforward to 
generalize the above approach in order to calculate dynamic correlation functions also for other operators. 

While we already introduced the time-dependent diffusion coefficient $D(t)$ for spin and energy transport in Eq.\ \eqref{Eq::Dt},  
the respective ac-conductivities at finite frequency $\omega$ are given by the Fourier transform of the current autocorrelations,  
\begin{align}
\text{Re}\ \sigma(\omega) &= \frac{1-e^{-\beta\omega}}{\omega L}\ \text{Re}\int\limits_0^\infty  
e^{i\omega t}\ \langle j_S(t) j_S \rangle\ \text{d}t\ , \label{Eq::SigOm} \\
\text{Re}\ \kappa(\omega) &= \beta \frac{1-e^{-\beta\omega}}{\omega L}\ \text{Re}\int\limits_0^\infty  
e^{i\omega t}\ \langle j_E(t) j_E \rangle\ \text{d}t\ \label{Eq::KappaOm}.
\end{align}
Note that the spin conductivity $\sigma(\omega)$ from Eq.\ \eqref{Eq::SigOm} must not be confused with the spatial variance $\sigma(t)$ introduced in Eq.\ \eqref{Eq::Sigma}. 
Omitting the possibility of finite Drude weights, the corresponding dc-conductivities are given by $\lim_{\omega\to 0} \sigma [\kappa](\omega) = \sigma [\kappa]_\text{dc}$ and 
can be connected to the diffusion constant via an
Einstein relation $D = D(t\to\infty) = \sigma[\kappa]_\text{dc}/\chi$, cf.\ Eq.\ \eqref{Eq::Dt}. Let us note that, since the Fourier transforms in Eqs.\ \eqref{Eq::SigOm} and \eqref{Eq::KappaOm}
can in practice be only evaluated up to a finite cutoff time $t_\text{max} < \infty$, 
the frequency resolution of $\sigma [\kappa](\omega)$ is finite as well [see also Eq.\ \eqref{Eq::FourierOm} in the main text]. 

\begin{figure}[tb]
 \centering 
 \includegraphics[width=0.8\columnwidth]{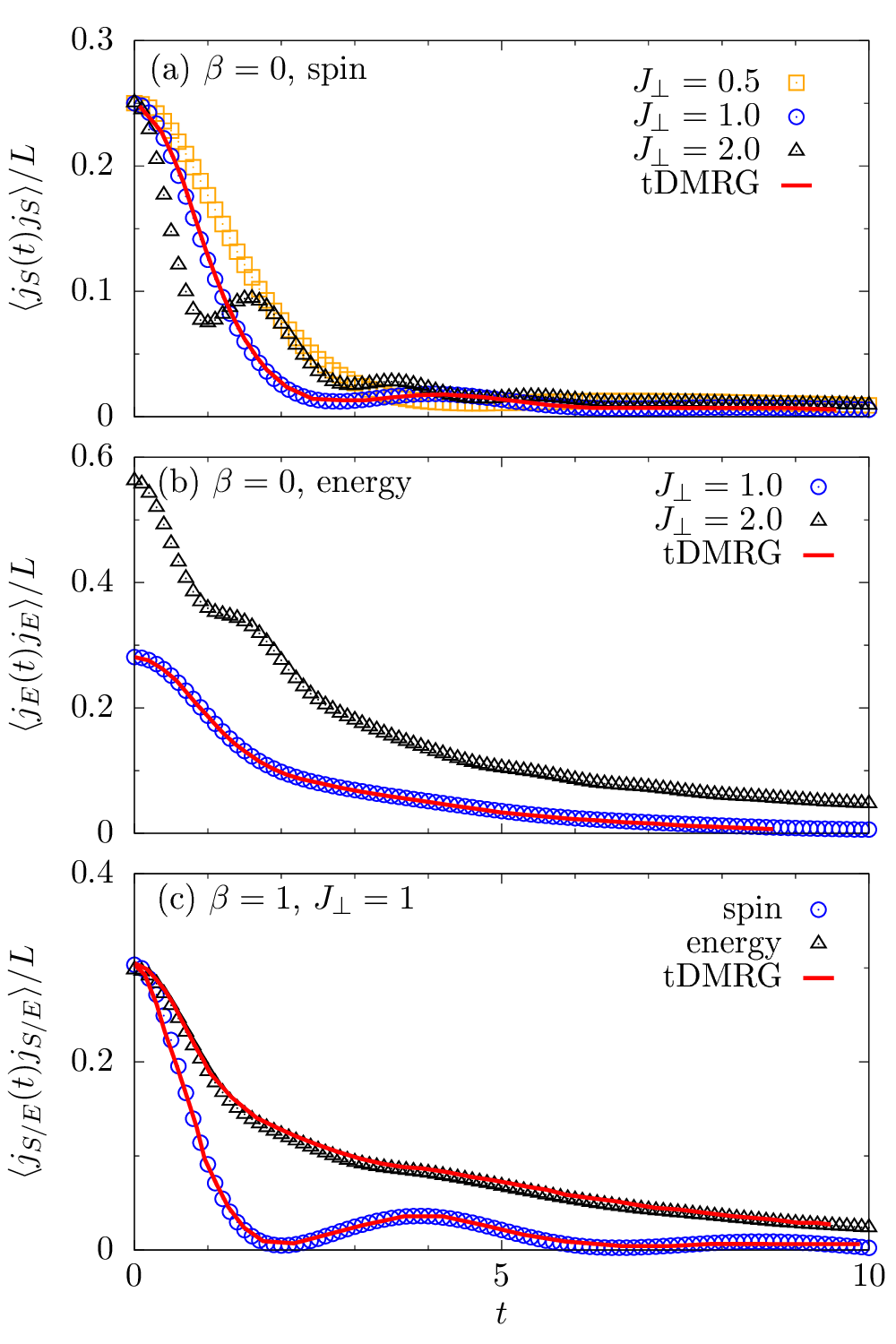}
 \caption{(Color online) Current autocorrelations. (a): Spin current $j_S$ for $J_\perp = 0.5, 1, 2$ and $\beta = 0$. 
 (b): Energy current $j_E$ for $J_\perp = 1, 2$ and $\beta = 0$. (c): Spin and energy current for 
 $\beta = 1$ and $J_\perp = 1$. For comparison, we depict tDMRG data digitized from Ref.\ \cite{Karrasch2015}.
 Note that in the case of $j_E$ we restrict ourselves to the symmetry subspaces with momentum $k = 0$.
 Other parameters: $J_\parallel = 1$, $L = 13$ (spin), $L = 15$ (energy).}
 \label{Fig10}
\end{figure}
In Figs.\ \ref{Fig10} (a) and (b), the current autocorrelations $\langle j(t) j\rangle$ at $\beta = 0$ are shown for spin and energy transport, 
respectively. While in Fig.\ \ref{Fig10} (a) we show data for smaller systems with $L = 13$ rungs only, the energy current in Fig.\ \ref{Fig10} (b) is calculated for systems 
with $L = 15$. Note however 
that in the latter case, we restrict ourselves to the symmetry subspaces with momentum $k = 0$ as the current is known 
to be essentially independent of $k$ for such system sizes. [This 
{\it crystal momentum} $k$ should not be confused with the wave number $k$ below Eq.\ \eqref{Eq::Fourier1}]. 
In all cases shown here, we observe that $\langle j(t) j\rangle$ decays to approximately zero, consistent 
with the absence of ballistic transport in a nonintegrable system. 
In the case $J_\perp/J_\parallel = 1$, our results are additionally compared to data digitized from Ref.\ \cite{Karrasch2015} obtained by a time-dependent density matrix renormalization group (tDMRG) approach. 
Generally, one finds a convincing agreement between both 
methods, i.e., our data for $L = 13, 15$ is free of significant finite-size effects and represents the thermodynamic limit. 
In Fig.\ \ref{Fig10} (c), spin and energy autocorrelations are depicted for the finite temperature $\beta = 1$. Also in this 
case, we observe that the pure-state method accurately reproduces the tDMRG data. Thus, typical pure states yield an 
efficient approach to correlation functions at finite temperatures as well. 

\begin{figure}[tb]
 \centering
 \includegraphics[width=0.8\columnwidth]{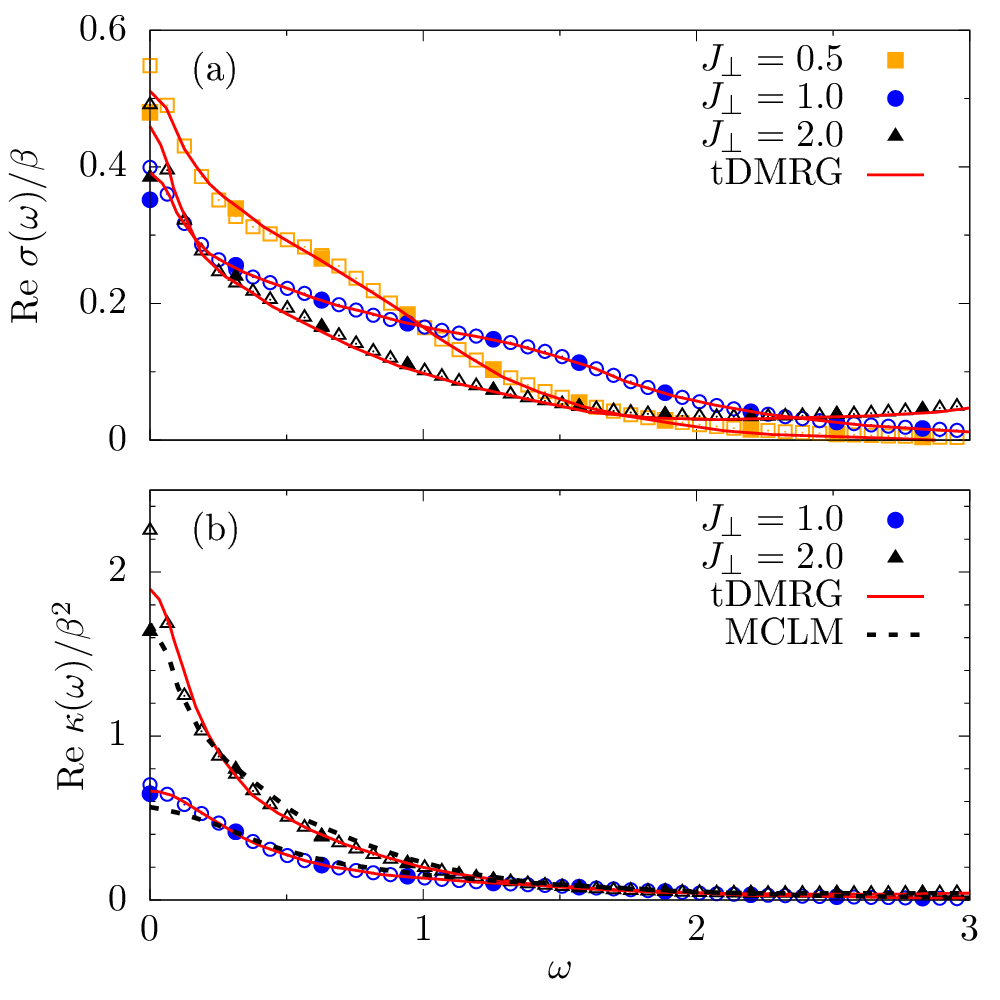}
 \caption{(Color online) (a): Spin conductivity $\sigma(\omega)$ for $J_\perp = 0.5, 1, 2$. (b): Energy conductivity $\kappa(\omega)$ for $J_\perp = 1, 2$. 
 We show data for two different frequency resolutions, $\delta \omega = \pi/10$ (filled symbols), $\delta \omega = \pi/50$ (open symbols). 
 For comparison, tDMRG data digitized from Ref.\ \cite{Karrasch2015} is shown. In (b), we also compare to data obtained from the microcanonical Lanzcos method (MCLM) \cite{Zotos2004}. 
 In the case of $\kappa(\omega)$ we restrict ourselves to the symmetry subspaces with momentum $k = 0$.
 Other parameters: $J_\parallel = 1$, $\beta = 0$, $L = 13$ (spin), and $L = 15$ (energy).}
 \label{Fig11}
\end{figure}

In Figs.\ \ref{Fig11} (a) and (b) the respective ac-conductivities $\sigma(\omega)$ and $\kappa(\omega)$ at $\beta = 0$ are 
shown, i.e., the Fourier transforms of the data in Figs.\ \ref{Fig10} (a) and (b). Particularly, we compare data with two different frequency resolutions
$\delta \omega = \pi/10$ and $\delta \omega = \pi/50$, i.e., a rather short and significantly longer cutoff time $t_\text{max}$ in Eqs.\ \eqref{Eq::SigOm} and \eqref{Eq::KappaOm}. 
In all cases, we observe a well-behaved dc-conductivity $\sigma[\kappa]_\text{dc} > 0$ which does not (significantly) depend on $t_\text{max}$, except $\kappa(\omega)$ for $J_\perp =2$. 
Moreover, our data is again in good agreement with existing data obtained by tDMRG \cite{Karrasch2015} and by 
the microcanonical Lanczos method \cite{Zotos2004}. 

\section{Trotter decomposition}\label{Sec::Trotter}

Let us briefly give some details on the time evolution of pure quantum states by means of a 
Trotter decomposition. 
To begin with, we note that the time-dependent Schr\"odinger equation
\begin{equation}
 i \partial_t \ket{\psi(t)} = \mathcal{H} \ket{\psi(t)} 
\end{equation}
is formally solved by
\begin{equation}\label{SolultionSE}
 \ket{\psi(t')} = U(t,t') \ket{\psi(t)}\ , 
\end{equation}
with $U(t,t') = e^{-i\mathcal{H}(t' -t)}$, where we have set $\hbar = 1$. 
While the exact evaluation of Eq.\ \eqref{SolultionSE} requires the diagonalization of $\mathcal{H}$, we here
approximate the time-evolution operator $U(t,t')$ by means of a Trotter product formula.
Specifically, a second-order approximation of $U(t,t+\delta t) = U(\delta t)$ is given by
\begin{equation}
\widetilde{U}_2(\delta t) = e^{-i \frac{\delta t}{2} {\cal H}_k} \cdots e^{-i 
\frac{\delta t}{2} {\cal H}_1} e^{-i\frac{\delta t}{2} {\cal H}_1} \cdots 
e^{-i\frac{\delta t}{2} {\cal H}_k} \, ,
\end{equation}
where ${\cal H}={\cal H}_1 + \cdots +{\cal H}_k$. This approximation is then bounded by
\begin{equation}
|| U(\delta t) - \widetilde{U}_2(\delta t) || \ll c_2 \, \delta t^3 \, ,
\end{equation}
where $c_2$ is a positive constant.
In practice, the Hamiltonian is decomposed into the 
$x$, $y$, and $z$ components of the spin operators, i.e., ${\cal H}={\cal H}_x 
+{\cal H}_y +{\cal H}_z$. Since the computational basis states are eigenstates of the 
$S^z$ operators, the representation $e^{-i \delta t {\cal H}_z}$ is 
diagonal by construction and only changes the input state by altering the 
phase of each of the basis vectors. Using an efficient basis rotation into the 
eigenstates of the $S^x$ or $S^y$ operators, the operators $e^{-i \delta t
{\cal H}_x}$ and $e^{-i \delta t {\cal H}_y}$ can act as $e^{-i \delta t 
{\cal H}_z}$ as well \cite{deReadt2006}.

\section{Fourier transforms}\label{Sec::FT}

Let us comment on the derivation of Eqs.\ \eqref{Eq::Fourier1} and \eqref{Eq::FourierOm} from the main part of this paper. 
Referring to Eq.\ \eqref{Eq::Typ}, we realize that a cut through the density profile $p_l(t)$ at fixed lattice site $l$ is equivalent to the correlation function 
$\langle \varrho_l(t) \varrho_{L/2}\rangle$. It is now instructive to perform the following calculation
\begin{align}
 \langle \varrho_q(t) \varrho_{-q}\rangle &= \frac{1}{L} \sum_{l=1}^L \sum_{l'=1}^L e^{iq(l-l')} \langle \varrho_{l}(t) \varrho_{l'} \rangle \\
					  &= \sum_{l=1}^L  e^{iql} \langle \varrho_{L/2+l}(t) \varrho_{L/2} \rangle \\
					  &= \sum_{l=1}^L  e^{iq(l-L/2)} p_l(t) = p_q(t)\ , 
\end{align}
where we have exploited translational invariance in order to compress the original double sum. Thus, we find that the 
intermediate structure factor $\langle \varrho_q(t) \varrho_{-q}\rangle$ can be easily obtained by a 
lattice Fourier transform of the real-space correlations. 
Furthermore, this momentum-space correlation function can also be transferred to what is usually referred to as the 
{\it dynamical structure factor} $S(q,\omega)$ by another Fourier transform from time to frequency domain, 
\begin{align}
S(q,\omega) &= \int_{-\infty}^\infty e^{i\omega t} \langle \varrho_q(t) \varrho_{-q}\rangle\ \text{d}t \\
	     &\approx \int_{-t_\text{max}}^{t_\text{max}}  e^{i\omega t} p_q(t)\ \text{d}t = p_q(\omega)\ ,  
\end{align}
where the finite cutoff time $t_\text{max}$ yields a frequency resolution $\delta \omega = \pi/t_\text{max}$. 
Thus, starting from the correlations $p_l(t)$ it 
is straightforward to also obtain correlation functions in momentum and frequency domain, 
which makes our pure-state method a rather powerful numerical approach. 

\section{Diffusion in lattice models}\label{Sec::Diff}

In this section, let us discuss in more detail how to detect diffusion in lattice models. 
To begin with, a process is called diffusive if it fulfills the lattice diffusion equation 
\begin{equation} \label{DiffEQ}
\frac{\text{d}}{\text{d}t}\ p_l(t) = D \left[ p_{l-1}(t) - 2 p_l(t) + p_{l+1}(t)\right]\ ,  
\end{equation}
with the \textit{time-independent} diffusion constant $D$. In the case of an initial $\delta$ peak profile at $l = L/2$, 
a specific solution for the time dependence of $p_t(t)$ is given by \cite{Richter2018_2}
\begin{equation}
p_l(t) - p_{\text{eq}} = \frac{1}{2} \exp(-2 D t) \, {\cal B}_{l-L/2}(2 D t)\ ,
\end{equation}
where ${\cal B}_l(t)$ is the modified Bessel function of the first kind and $p_\text{eq} = p_{l\neq L/2}(0)$ denotes the homogeneous background. (Note that in this 
paper we have $p_\text{eq} = 0$ since $\text{Tr}[\varrho_l] = 0$.) 
In case of  
sufficiently large $L$ and long $t$, i.e., if the discrete lattice momenta $q$ become 
dense, this lattice solution can be well approximated by a Gaussian. Such Gaussians have been observed in Figs.\ \ref{Fig3} (a) and (b).
Specifically, the spatial variance of these Gaussians is then also given by $\sigma^2(t) = 2Dt$, i.e., $\sigma(t) \propto \sqrt{t}$. 

Given some general density distribution, it is in some cases instructive to study the dynamics in momentum space as well. 
In this context, a Fourier transform of Eq.\ \eqref{DiffEQ} yields 
\begin{align}\label{DiffMom}
\frac{\text{d}}{\text{d}t}\ p_q(t) = -2D(1-\cos q) p_q(t)\ .
\end{align}
Apparently the $L$ different Fourier modes $p_q(t)$ are completely decoupled and  
their exponentially decaying solutions have been already given in Eq.\ \eqref{Eq::ExpLor}. 



\begin{thebibliography}{99}
\bibitem{Suscep}
For $\varrho_l = S_{l,1}^z + S_{l,2}^z$ we have $\chi = 1/2$, while for $\varrho_l = h_l$ we find $\chi = 3(2J_\parallel^2 + J_\perp^2)/16$.
\bibitem{InitDelta}
Note that in the case of energy dynamics, we also find two smaller peaks at adjacent lattice sites $L/2 \pm 1$. 
This is caused by the fact that $[h_l,h_{l\pm1}] \neq 0$.
\end{thebibliography}
\end{document}